\documentclass[12pt,fleqn]{article}
\usepackage[english]{babel}
\begin{document}

\begin{center}
{
\large
{\bf Exact Travelling Wave Solutions of Some Nonlinear Nonlocal Evolutionary
Equations}
}

\vspace{5mm}
 {\bf Vsevolod A. Vladimirov
and Ekaterina V. Kutafina} \\
University of Science and Technology\\
Faculty of Applied Mathematics \\ Al Mickiewicza 30,
 30-059 Krak\'{o}w, Poland \\ E-mail: vladimir@mat.agh.edu.pl, vsan@rambler.ru
\vspace{5mm}

\end{center}

{
\footnotesize
\noindent
{\bf Abstract.}
Direct algebraic method of obtaining exact solutions to nonlinear
PDE's is applied to certain set of nonlinear nonlocal evolutionary equations,
including nonlinear telegraph equation, hyperbolic generalization of  Burgers
equation and some spatially nonlocal hydrodynamic-type model.
Special attention is paid to the construction of the kink-like and
soliton-like solutions.
}

\section{ Introduction }

In recent decades the problem of obtaining exact solutions of
nonlinear evolutionary equations has attracted attention of many
experts in mathematical physics. The most fundamental achievement in this
area was development of the inverse scattering method \cite{dodd}.
Unfortunately, this method is applicable to the relatively narrow
class of the completely integrable equations, while for the
majority of nonlinear evolutionary equations methods of obtaining
the general solution do not exist. Yet even the possibility of
obtaining particular solutions to the nonlinear PDE's very often
is considered  in applications as a big success, because having
the analytical solutions of a modelling system it is more easy to
analyze it and to make its interpretation. Exact solutions are
also widely used as a starting point for various asymptotic
methods, for testing the numerical schemes and
facilitating the stability analysis.

For nonlinear PDE's, which are not integrable, exact solutions are
usually obtained by means of the Group Theory Reduction
\cite{olv}. Appreciating all advantages of the symmetry-based
methods, we would like to pay attention to the fact that they are
not fully universal, firstly, because they can be effectively used
merely in case the  PDE's possess some non-trivial symmetry,
and, secondly, because  it is very difficult within the frameworks
of these methods  to obtain any solution with the given
properties. Note that much more efficient from this point of view
is the combination of  the symmetry reduction with the methods of
qualitative analysis.

Aside from the symmetry reduction, there exist another group of
methods enabling to obtain  exact solutions to nonlinear PDE's.
They are based on choosing the proper transformation (or ansatz),
simplifying the problem. As an example let us mention the famous
Cole-Hopf transformation, which was originally used for the
non-local linearization of Burgers equation. Later on employment
of the similar transformations enabled  Hirota to obtain the
multi-soliton solutions of the completely integrable KdV equation
without referring to the inverse scattering method  \cite{dodd}.
An intense development of the ansatz-based method in the following
years was highly simulated by the fact that it proved  to be
effective for obtaining the particular exact solutions  of the
evolutionary PDE's that are not completely integrable. The essence
of the ansatz-based method is well summarized in the recent work
of E. Fan \cite{fan}. Beside the imposing bibliography, it
contains general formulation of the unified algebraic method,
which will be presented in next section. Literally during last
years the approach presented by E.Fan  was used in the composition
with the generalized  Cole-Hopf anzatz, which made possible to
obtain a series of exact solutions of nonlinear transport equation
\cite{nikbar,baryur}.

In this paper we present certain modification of the anzatz-based
method presented in \cite{nikbar}. In accordance with the core of
our interests, we use it to obtain exact solutions of hyperbolic
modification of the non-linear transport equation, which takes
into account the nonlocal effects.  We concentrate on finding out
the particular types of travelling wave (TW) solutions, namely,
the soliton-like and kink-like solutions, but this approach can be
easily used for searching out any exact solution which can be
described as an algebraic combination of certain types of  special
functions. Next we consider the family of TW solutions for the
spatially nonlocal hydrodynamic-type  model. Using the qualitative
analysis we state the existence of periodic and soliton-like TW
solutions. Imposing some restrictions on the parameters we obtain
the soliton-like solution in the analytical form. It turns out to
be much more complicated than that obtained within the above
mentioned  anzatz-based methods.

\section{Exact TW solutions to the hyperbolic modification of nonlinear
transport equations}

\subsection{Unified algebraic method and it's modifications}

The essence of the anzatz-based method or the {\it direct  unified
algebraic method} in E.Fan's terminology \cite{fan}, is based on
the observation that the particular solutions of any system of
PDE's
\begin{equation}\label{genevol}
H^\nu(u^i,\,\,u^i_t,\,u^i_x,\,u^i_{xx},...)\qquad i=1,2,...m,
\quad \nu =1,2,..,n
\end{equation}
which do not depend on $(t,x)$ coordinates in explicit form,  can
be presented as a linear  combination
\begin{equation}\label{genansatz}
u^i=\sum_{\mu=0}^{n}a_\mu^i \phi^\mu(\xi) \qquad \xi=x+vt,
\end{equation}
where $a_\mu$ are unknown parameter, while function $\phi(\xi)$ -
satisfies the equation
\begin{equation}\label{phieqn}
\phi(\xi)=\pm \sqrt{\sum_{\nu=0}^r c_\nu \phi^\nu}.
\end{equation}
Depending on the conditions posed on  the parameters $c_\nu,$
solutions of the equation (\ref{phieqn}) are expressed by the
elliptic Jacobi (Weierstrass), hyperbolic or trigonometric
functions \cite{fan}. The properties of these functions are
inherited by those solution  of the initial system that can be
presented in the form (\ref{genansatz}).
In fact this methodology is constructive and  algorythmic  if the
functions $H^{\nu}$ arising in (\ref{genevol}) are algebraic ones.
One easily gets convinced that with this assumptions the
substitution of (\ref{genansatz}) into the initial system
(\ref{genevol}) gives the polynomial functions with respect to
$\phi^\mu(\xi)$ and $\phi^\mu\,\dot\phi(\xi)$. Equating to zero
the coefficients standing at the corresponding powers of these
functions, we obtain nonlinear system of the algebraic equations
which determines particular solutions of the initial system.

The enforced version of the Fan's method was  put forward recently
in \cite{nikbar,baryur}. It was used for searching out the exact
solutions of the equation
\begin{equation}\label{baniy}
u_t+Auu_x-\kappa u_{xx}=f(u).
\end{equation}
More preciesly, there has been  proposed the ansatz
\begin{equation}\label{ansnik}
u=\left[\frac{z'(\xi)}{z(\xi)}\right]^k, \qquad \xi=x+vt+x_0,
\qquad k=1,2,..,
\end{equation}
where $z=\sum a_\mu \phi^\mu(\xi)$, and $\phi(\xi)$ is the
function satisfying the  (\ref{phieqn}). Owing to this
combination, the multi-parameter families of the exact solutions
were obtained in the situation when the pure Fan's methodology
does not work \cite{nikbar}.

On analyzing different versions of the anzatz-based method, one can conclude
that effectiveness of their employment is based on the mere fact
that the family of functions $\phi^\mu(\xi)$ and
$\phi^\nu\dot{\phi }(\xi)$ is closed with respect to the algebraic
operations and differentiating. In view of this a quite natural
generalization of the already mention  ansatze would be as
follows:
\begin{equation}\label{vkgenans}
 u=\frac{f(\xi)}{g(\xi)}=\frac{\sum_{\mu=0}^{m_1} a_\mu \phi^\mu(\xi)+
 \dot{\phi(\xi)}\sum_{\nu=0}^{m_2} b_\nu \phi^\nu(\xi)}
 {\sum_{\lambda=0}^{n_1} c_\lambda \phi^\lambda(\xi)+
 \dot{\phi(\xi)}\sum_{\kappa=0}^{n_2} d_\kappa \phi^\kappa(\xi)},
 \end{equation}
where  the function $\phi(\xi)$   still satisfies the equation
(\ref{phieqn}), but, in contrast to (\ref{ansnik}), dependence
between the functions $f$ and $g$ is not assumed from the very
beginning. Effectiveness of the anzatz (\ref{vkgenans}) is
demonstrated in the following subsection.

\subsection{Exact travelling wave solution to the nonlinear hyperbolic equation}

Let us consider the following equation:
\begin{equation}\label{vkhypeq}
\tau u_{tt}+Auu_x+Bu_t-\kappa u_{xx}=f(u)=\sum_{\nu \in
I}\lambda_\nu u^\nu,
\end{equation}
where $\tau$, $A$, $B$, $\kappa$ are non-negative constants.
 For $A=0$ equation (\ref{vkhypeq}) coincides with the
nonlinear telegraph equation; for $A\ne 0$ it
coincides with the hyperbolic generalization of  Burgers equation,
while for $A=B=0$ -- with the nonlinear d'Alambert
equation.The hyperbolic modifications of nonlinear
transport equations arise in a natural way when the memory effects
are taken into account \cite{makar1}.

 The main goals of this paper are to obtain the exact soliton-like
 and kink-like solutions of equation (\ref{vkhypeq}) and  to present
 the advantages of anzatz (\ref{vkgenans}) compared to the already
 mentioned modification \cite{nikbar} of the unified algebraic method.

Assuming that the solitons and kinks can be expressed by powers of
function $sech(\xi)$, which is the particular solution of equation
(\ref{phieqn}) and function $sinh(\xi)$ which appears in the odd
derivatives of the function $sech(\xi)$, we use the following
ansatz:
\begin{equation}\label{shans}
u(\xi)=\frac{f(\xi)}{g(\xi)}=\frac{\sum_{\mu=0}^{m_1}a_\mu sech^\mu
(\alpha\xi)+ sh(\alpha\xi)\sum_{\nu=0}^{m_2}b_\nu sech^\nu
(\alpha\xi)} {\sum_{\gamma=0}^{n_1}c_\gamma sech^\gamma
(\alpha\xi)+sh(\alpha\xi) \sum_{\sigma=0}^{n_2}d_\sigma
sech^\sigma (\alpha\xi)}
\end{equation}
or, what is the same,
\begin{equation}\label{expans}
u(\xi)=\frac{\sum_{\mu=0}^{m}a_\mu exp(\mu\alpha\xi)}
{\sum_{\nu=0}^{n}b_\nu exp(\nu\alpha\xi)}.
\end{equation}

Inserting  anzatz (\ref{shans}) ( or (\ref{expans})) into (\ref{vkhypeq}) and executing
of all necessary operations, we obtain an algebraic equation
containing, respectively, functions $sech^\mu(\alpha\xi)$,
$sech^\nu(\alpha\xi)sh(\alpha\xi)$ or $\exp{[\mu\,\alpha\,\xi]}$.
Regarding them as the functionally independent ones
and equating to zero corresponding coefficients, we go to the
system of algebraic equations. We do not expose the details of
these calculations since they are simple but cumbersome. To
accomplish them we  used the package of symbolic computation
"Mathematica". The results obtained are presented below.

%%%%%%%%%% Begin

%\noindent
I. For arbitrary $A,\,\, B$ and
%\begin{itemize}
%\item
%[Stare rozwiazania tego typu: 1. plik VK1E - typu tgh; 2. plik VK2E (3.1); 3.VK3E -drugie rozwiazanie; 4.VK4E na gorze kartki.]
$f(u)=\lambda_0+\lambda_1 u(t,x)+\lambda_2 u(t,x)^2+\lambda_3 u(t,x)^3$ function
\begin{equation}\label{vk1_1}
u(t,x)=\frac{a_0+a_1e^{ \alpha (x+vt)}}{b_0+b_1e^{ \alpha (x+vt)}}
\end{equation}
satisfies equation (\ref{vkhypeq}) if the following conditions
hold:
%\begin{itemize}
%\item
%[Stare rozwiazania tego typu: 1. plik VK1E - typu tgh;
%2. plik VK2E (3.1); 3.VK3E -drugie rozwiazanie; 4.VK4E na gorze kartki.]
{
%\footnotesize
\begin{equation}\label{vk1_1aux}
\begin{array}{ll}
\lambda_0=\frac{-a_0a_1\alpha}{ \Delta^2 }(Bv\Delta+h\Theta), \quad
\lambda_1=\frac{1}{b_0b_1\Delta^2}\left[\alpha b_0b_1(Bv\Theta\Delta+h\Theta^2)+\lambda_3a_0a_1\Delta^2\right], \\
\\
\lambda_2=\frac{-1}{b_0b_1\Delta^2}\left[\alpha
b_0^2b_1^2(Bv\Delta+h\Theta)+\lambda_3\Delta^2\Theta\right],
\quad A=\frac{-1}{\alpha b_0b_1\Delta}\left[-2h\alpha
b_0^2b_1^2+\lambda_3\Delta^2\right].
 \end{array}
\end{equation}
} Here and henceforth we use the notation $
h=\alpha(v^2\tau-\kappa), \quad \Delta=a_1b_0-a_0b_1, \quad
\Theta=a_1b_0+a_0b_1.$ Equation (\ref{vk1_1}) defines a kink-like
regime when $b_0 b_2>0$ and ${a_0}/{b_0}\neq{a_2}/{b_2}$. Using
the conditions (\ref{vk1_1aux}), we can express the unknown
parameters from formula (\ref{vk1_1}) by the parameters
characterizing equation (\ref{vkhypeq}), yet, in general case it
is too cumbersome. It is much easy to do when $\alpha=2\sqrt{-\lambda_0\lambda_2}/v$, $a_0=-a_1=\sqrt{-\lambda_0/\lambda_2}$, $b_0=b_1=1$. From these conditions we obtain the solution:
\[
u(t,\,x)=\sqrt{-\lambda_0/\lambda_2}Tanh\left[\frac{\sqrt{-\lambda_0\lambda_2}}{v}(x+vt)\right]
\]
and conditions:
 $v=\frac{\lambda_2(A\,B+\sqrt{A^2\,B^2-8\kappa\,\lambda_3+16\kappa\,\lambda_2^2\,\tau})}{2\lambda_3-4\lambda_2^2\,\tau}$, $\lambda_1=\frac{\lambda_0\,\lambda_3}{\lambda_2} $.
For  $\kappa=1$, $\tau=0$, $A=0$, $B=1$
function (\ref{vk1_1}) coincides with the solution obtained in
\cite{nikbar}.

Another example of the kink-like solution defined by the formulae
(\ref{vk1_1})-(\ref{vk1_1aux}) is as follows:
\[
u(t,\,x)=\frac{2}{b_0[1+\exp({2\,\alpha\,\xi})]}, \qquad \xi=x+vt,
\]
where $
b_0={(-\lambda_2\pm\sqrt{\lambda_2^2-4\,\lambda_1\,\lambda_3})}/{\lambda_1}$, $A=\lambda_0=0$,
$\alpha=-{(2\lambda_2+3\,b_0\,\lambda_1)}/{[4B\,v\,b_0]}$, and
\[
v=\frac{\pm\sqrt{\kappa}(2\lambda_2+3\,b_0\,\lambda_1)}{\sqrt{4\lambda_2^2\tau+4b_0\lambda_2(B^2+3\lambda_1\tau)+b_0^2\lambda_1(2B^2+9\lambda_1\tau)}}.
\]
% or $$\sqrt{-\lambda_0/\lambda_2}Tanh^{-1}(\frac{\sqrt{-\lambda_0\lambda_2}}{\mu}(x+vt)).$$

II. For $A=0$, arbitrary $B$  and
$f(u)=\lambda_0+\lambda_{1/2}u(t,x)^{\frac{1}{2}}+\lambda_1
u(t,x)+\lambda_{3/2} u(t,x)^{\frac{3}{2}}+\lambda_2 u(t,x)^2$
function
\begin{equation}\label{vk2}
u(t,x)=\left[\frac{a_0+a_1e^{ \alpha (x+vt)}}{b_0+b_1e^{ \alpha (x+vt)}} \right]^2
\end{equation}
satisfies  the equation(\ref{vkhypeq}) providing that the
following conditions hold:
%warunki
{
%\footnotesize
\[ \begin{array}{ll}
\lambda_0={2a_0^2a_1^2\alpha h}/{\Delta^2}, \quad  \lambda_{1/2}={-2a_0a_1\alpha}(3h\Theta+Bv\Delta)/{\Delta^2},\\
\\
\lambda_1={2\alpha}(h(3\Theta^2-\Delta^2)+Bv\Delta\Theta)/{\Delta^2}, \quad \lambda_{3/2}={-2b_0b_1\alpha}(5h\Theta+Bv\Delta)/{\Delta^2},\\
\\
\lambda_2={6b_0^2b_1^2h\alpha}/{\Delta^2}.
 \end{array}
 \]
 }
This solution defines the solitary wave regime if $b_0 b_1>0\,$,
${|a_0|}/{|b_0|}={|a_2|}/{|b_2|}\,$ while ${a_0}/{b_0}\neq
{a_2}/{b_2}.$
%h=\alpha(v^2\tau-\kappa) &
%\Delta=a_1b_0-a_0b_1& \Theta=a_1b_0+a_0b_1

III. For $B=0$, arbitrary $A$  and $f(u)= \lambda_1 u(t,x)+\lambda_3 u(t,x)$
function
\begin{equation}\label{vk3}
u(t,x)=\frac{a_1e^{\alpha \xi}+a_2e^{2\alpha
\xi}}{-a_1^3-3\,a_1^2\,a_2\,e^{\alpha
\xi}+3\,a_1\,a_2^2\,e^{2\alpha \xi}+a_3^3\,e^{3\alpha \xi}},
\qquad \xi=x+vt
\end{equation}
satisfies (\ref{vkhypeq}), when $\lambda_1,\,\lambda_3$ are
positive and the parameters are as follows: $\,\,
6\,a_1\,a_2=-\sqrt{\lambda_3/\lambda_1}$,
$\alpha={\sqrt{\lambda_1\,\lambda_3}}/{A}$,
$\,\,v=\pm\sqrt{\left(A^2/\lambda_3+\kappa\right)/{\tau}}.\,\,$
This solution is always singular, because for arbitrary values of
the parameters the expression in the denominator of the formula
(\ref{vk3}) nullifies  for some  $\xi\in R^1.$

IV. Now let us consider the case ${A=B=0.}$

\noindent IV a. For $f(u)=\lambda_0+\lambda_1 u(t,x)+\lambda_2
u(t,x)^2+\lambda_3 u(t,x)^3$
 function
\begin{equation}\label{vk4_1}
u(t,x)=\frac{a_0+2a_1e^{\alpha (x+vt)}+a_0e^{2\alpha (x+vt)}}{b_0+2b_1e^{\alpha (x+vt)}+b_0e^{2\alpha (x+vt)}},
\end{equation}
%warunki:
satisfies equation (\ref{vkhypeq}) when the following conditions
hold: {
%\footnotesize
\begin{equation}\label{vk4_1aux}
\begin{array}{lll}
\lambda_0=a_0\,(2a_0^2\,b_0-a_1^2\,b_0-a_0\,a_1\,b_1\,)\alpha\,h/\Delta^2\\
\lambda_1=((a_1^2\,b_0^2+4a_0\,a_1\,b_0\,b_1+a_0^2\,(-6\,b_0^2+b_1^2))\alpha\,h/\Delta^2\\
\lambda_2=3\,b_0\,(2\,a_0\,b_0^2-a_1\,b_0\,b_1-a_0\,b_1^2)\alpha\,h/\Delta^2\\
\lambda_3=-2\,b_0\,(b_0^2-b_1^2)\alpha\,h/\Delta^2
 \end{array}
\end{equation}
} For $a_0\neq 0$, $b_0 \neq 0$, $|a_1|+|b_1|\neq 0$ equation
(\ref{vk4_1}) defines  the soliton-like solution. One of the
parameters, contained in (\ref{vk4_1}) can be chosen arbitrarily,
while the rest can be expressed, using the (\ref{vk4_1aux}), as
the functions of the parameters, defining equation
(\ref{vkhypeq}). We omit doing this in the general case, but
present one particular example. Thus, for $b_1=0,\,\,b_0=1$ and
arbitrary $\alpha$ we have the solution
\[
u(t,\,x)=\frac{a_0+2a_1e^{\alpha (x+vt)}+a_0e^{2\alpha (x+vt)}}{1+e^{2\alpha (x+vt)}},
\]
with $\,\, a_0=-{\lambda_2}/{(3\,\lambda_3)}$,
$\,\,a_1=\sqrt{2{(\lambda_2^2-\lambda_1\,\lambda_3)}/{\lambda_3^2}}$,
$\,\,
v=\pm\sqrt{\left[\lambda_1-{\lambda_2^2}/{(3\,\lambda_3)}+\kappa\,\alpha^2
\right]/({\tau\,\alpha^2})}$, with the additional condition $\lambda_0+\lambda_1a_0+\lambda_2a_0^2+\lambda_3a_0^3=0$.

%\item[(b)]
%[VK6E od gory]\\
\noindent VI b. For $f(u)=\lambda_{1/2}
u(t,x)^{{1}/{2}}+\lambda_1 u(t,x)+\lambda_{3/2}
u(t,x)^{{3}/{2}}+\lambda_2 u(t,x)^{2}$ we obtain the
soliton-like solution
\begin{equation}\label{vk4_2}
u(t,x)=\frac{(e^{\alpha (x+vt)}+1)^4}{(b_0e^{2\alpha
(x+vt)}+(2b_0+4b_1)e^{\alpha (x+vt)}+b_0)^2}
\end{equation}
with
%warunki:
\[\begin{array}{lll}\lambda_{1/2}=-3\,\alpha\, h/b_1 & \lambda_1=(12\,b_0+4\,b_1)\alpha \,h/b_1 \\
\lambda_{3/2}=-(15\,b_0^2+10\,b_0\,b_1)\alpha\, h /b_1 & \lambda_2=(6\,b_0^2\,b_1+6\,b_0^3)\alpha\, h/b_1.\end{array}\]

%
%\item[(c)]
%[VK6E drugie rozwiazanie]
\noindent
VI c. For
$f(u)=\lambda_0+\lambda_{1/2} u(t,x)^{\frac{1}{2}}+\lambda_1 u(t,x)+\lambda_{3/2} u(t,x)^{\frac{3}{2}}$
the localized wave pack
\begin{equation}\label{vk4_3}
u(t,\,x)=\frac{(a_0e^{2\alpha (x+vt)}+(2a_0+4a_1)e^{\alpha
(x+vt)}+a_0)^2}{(e^{\alpha (x+vt)}+1)^4}
\end{equation}
%warunki:
defines a solution of (\ref{vkhypeq}) if the following conditions
hold:
\[\begin{array}{lll}
\lambda_0=2a_0^2(a_0+a_1)\,\alpha \,h/a_1 &
 \lambda_{1/2}=(9a_0^2+6a_0a_1)\,\alpha \,h/a_1 \\
\lambda_{1}=(12a_0+4a_1)\,\alpha \,h/a_1&
 \lambda_{3/2}=-5\,\alpha \,h/a_1

.\end{array}\]

%\item[(d)]
%[VK6E trzecie rozwiazanie]
\noindent VI d. For $f(u)=\lambda_1 u(t,x)+\lambda_{3/2}
u(t,x)^{\frac{3}{2}}+\lambda_2u(t,x)^2$ function
\begin{equation}\label{vk4_4}
u(t,x)=\frac{4e^{2\alpha (x+vt)}}{(a_0e^{2\alpha
(x+vt)}+2a_1e^{\alpha (x+vt)}+a_0)^2}
\end{equation}
with \[\begin{array}{lll} \lambda_1=4\,\alpha \,h &
\lambda_{3/2}=-10\,a_1\alpha\, h \\
 \lambda_2=(6\,a_1^2-6\,a_0^2)\alpha\, h.\end{array}\]
  defines the soliton-like solution of the
equation (\ref{vkhypeq}).
%
%\item[(e)]

\noindent VI e. Finally, let us consider equation
\begin{equation}\label{dalamb}
\tau u_{tt}-\kappa u_{xx}=\lambda_0+\lambda_1
u+\lambda_2u^2+\lambda_3u^3.
\end{equation}
Inserting the anzatz $u=\phi(\xi), \quad \xi=x+v\,t$ into the equation
(\ref{dalamb}), we obtain, after one integration, the following
ODE:
\begin{equation}\label{fans}
\frac{d\phi}{d\xi}=\pm\,\sqrt{c_0+c_1\,u+c_2\,u^2+c_3\,u^3+c_4\,u^4},
\end{equation}
where $c_0$ is an arbitrary constant, $ c_1=2\lambda_0/H$,
$c_2=\lambda_1/H$, $ c_3=2\lambda_2/(3\,H)$,
$c_4=\lambda_3/(2\,H),$ $H=\tau\,v^2-\kappa.$  To this equation
the classification given in \cite{fan} is applied:
\begin{itemize}
\item[(a)]
 if $\lambda_2=\lambda_0=0$, then equation (\ref{dalamb}) possesses a soliton-like solution
\begin{equation}\label{vk4_5a}
u(t,\,x)=\sqrt{-2\lambda_1/\lambda_3}sech(\sqrt{\lambda_1/H}\xi)\qquad\lambda_1>0,\qquad
\lambda_3<0;
\end{equation}
 \item[(b)] if $\lambda_0=\lambda_2=0$,
 then equation (\ref{dalamb}) posesses a kink-like solution
\begin{equation}\label{vk4_5b}
u(t,\,x)=\sqrt{-\lambda_1/\lambda_3}\,\tanh{\left[\sqrt{-\lambda_1}/(2\,H)\,\xi\right]},\qquad\lambda_1<0,\qquad
\lambda_3>0;
\end{equation}
 \item[(c)] if $\lambda_0=\lambda_3=0$, then
equation (\ref{dalamb}) possesses a soliton-like solution
\begin{equation}\label{vk4_5c}
u(t,\,x)=-\left[3\lambda_1/(2\lambda_2)\right]{sech}^2{\left[\sqrt{\lambda_1/H}\,\xi/2\right]}\qquad\lambda_1>0.
\end{equation}
 \end{itemize}
%\end{enumerate}

%%%%%%%%%%%% End

Presented above results enable us to state that the anzatze
(\ref{shans}) and (\ref{expans}) are effective and their
employment gives the exact solutions in the situations when
the ansatze suggested in \cite{nikbar,fan} do not work. Note,
that as a by-product we obtained  a number of new exact solutions
of non-linear transport and Burgers equations. These solutions can
be easily extracted from the presented above formulae by simple
substitution $B=1,\tau=0$ (and also $A=0$ when it is necessary).
Thus, the proposed modification of the unified algebraic method
proves to be useful and its employment results in essential
broadening the number of  solutions of the given type (i.e.
soliton-like and kink-like solutions), which can be obtained in
analytic form. Yet, as it will be shown in the following section,
none of the version of the anzatz-based method proposed by now is
fully universal.

\section{Periodic and soliton-like TW solutions of the nonlocal
hydrodynamic-type model.}

In conclusion, let's analyze the  family of TW solutions for
the following system:
\begin{equation}\label{hdyngen}
\begin{array}{l}
 u_t+\beta \rho^{\nu+1}\rho_x+\sigma \left[\rho^{\nu+1}\rho_{xxx}+3(\nu+1)\rho^{\nu}
\rho_{x}\rho_{xx}+\nu(\nu+1)\rho^{\nu+1}\rho^{\nu-1}\rho_{x}^3\right]=0, \\
\rho_{t}+\rho^{2}u_x=0,
\end{array}
 \end{equation}
where $\nu,\beta,\sigma$ are constants. The system
(\ref{hdyngen}) arises in a natural when the balance equations
for mass and momentum, taken in the hydrodynamic approximation,
are closed by the dynamic equation of state, accounting for
the short-ranged spatial non-locality \cite{vsan2}.
In general case the answer on the existence of  the periodic
and soliton-like  TW solutions is obtained by the methods of
qualitative analysis, but under some additional conditions posed
on the parameters we are able to present some of them in the
analytic form, omitting the anzatz-based method (which doesn't
work in this case).

Let us consider the following family of invariant travelling wave
solutions:
\begin{equation}\label{anspat}
u=U(\omega),\qquad \rho=R(\omega), \qquad \omega=x-Dt.
\end{equation}
Inserting the anzatz (\ref{anspat}) into the second equation of
system (\ref{hdyngen}), we obtain the first integral $
U=C_{1}-{D}/{R} $ and the system of ODE's
\begin{equation}\label{ds1sp}
\left\{
\begin{array}{l}
\frac{dR}{d\omega}=Y \\
\frac{dY}{d\omega}= \left(\sigma R^{\nu+2}\right)^{-1}
\left\{ER-\left[D^{2}+{\beta}R^{\nu+3}/{(\nu+2)}+
\sigma(\nu+1)R^{\nu+1}Y^{2}\right]\right\}.
\end{array}
\right.
\end{equation}
In accordance with the asymptotic conditions $ \lim_{\omega\to
+\infty}U(\omega)=0, \quad\lim_{\omega\to
+\infty}R(\omega)=R_{1}>0,$ $U(+\infty)=0,\,\,\,R(+\infty)=R_1>0,$
we assume henceforth that $ C_1=D/R_1$ and
$E={D^{2}}/{R_{1}}+{\beta}\,R_{1}^{\nu+2}/{(\nu+2)}$.

Dividing the second equation of system (\ref{ds1sp}) by the first
one and introducing new variable $Z=Y^2\equiv (d\,R/d\,\omega)^2,$
we get, after some algebraic manipulation, the linear
inhomogeneous equation
\begin{equation}\label{lininhom}
{Z'}{(R)}+{2}{[(\nu+1)\,R]^{-1}}Z(R)={2}
\left[E\,R-D^2-{\beta}R^{\nu+3}/{(\nu+2)}
\right]/(\sigma\,R^{\nu+2}).
\end{equation}
Solving this equation with respect to $Z=Z(R)$ and next
integrating the equation obtained after the substitution
$Z=(d\,R/d\,\omega)^2$, we can express the solution of
(\ref{ds1sp}) as the following quadrature:
\begin{equation}\label{solds1spat}
\omega-\omega_0=\int\frac{\pm\sqrt{\sigma}\,R^{1+\nu}\,d\,R}{\sqrt{H_1+2\,E\,\frac{R^{2+\nu}}{(2+\nu)}-
2\,D^2\frac{R^{1+\nu}}{(1+\nu)}-\beta\,\frac{R^{2(2+\nu)}}{(2+\nu)^2}}}.
\end{equation}

Unfortunately, the direct analysis of  solution (\ref{solds1spat})
is  very difficult. To realize what sort of solutions we deal
with, the qualitative analysis of system (\ref{solds1spat}) is
performed. It is evident, that all isolated critical points of
system (\ref{ds1sp}) are located on the (horizontal) axis  $OR$.
They are determined by  solutions of the algebraic equation $
P(R)={\beta}\,R^{\nu+3}/{(\nu+2)}-ER+D^{2}=0.$
One of the roots of this equation coincides with $R_{1}$.
Location of the second real root depends on relations between the
parameters. If  $\nu>-2$ and $D^2>\beta R^{\nu+3}_{1}$, then
there exists the second critical point $A_2(R_2,\,0)$ with
$R_{2}>R_{1}$ and  the polynomial $P(R)$ has the  representation
\begin{equation}\label{p_repr}
P(R)=(R-R_{1})(R-R_{2})\Psi (R).
\end{equation}
For any $\nu >-2$ function  $\Psi (R)$ is positive whenever $R>0$,
since the function $\beta\,R^{\nu+3}/(\nu+2)$ is concave
in this interval and has exactly two intersections with the line
$E\,R-D^2.$

Analysis of system's (\ref{ds1sp}) linearization matrix shows,
that the critical points $A_{1}(R_{1},0)$ is a saddle, while the
critical point  $A_{2}(R_{2},0)$ is a center.  Thus,  system
(\ref{ds1sp}) has only such critical points,  that are
characteristic to the hamiltonian system. This circumstance
suggests that there could exist a hamiltonian system equivalent to
(\ref{ds1sp}) In fact, the following statement holds.

{\bf Lemma.} {\it If to introduce a new independent variable $T,$
obeying the equation $ {d\,\omega}/{dT}=2\,\sigma
R^{2(\nu+1)},$ then system (\ref{ds1sp}) can be
written as a hamiltonian one
\begin{equation}\label{ds1h}
\left\{
\begin{array}{ll}
 \frac{dR}{dT}=2\,\sigma R^{2(\nu+1)}\equiv\frac{\partial H}{\partial Y}, \\
\frac{dY}{dT}= 2\,R^{\nu}
\left(ER-\left[D^{2}+{\beta}\,R^{\nu+3}/{(\nu+2)}+\sigma(\nu+1)R^{\nu+1}Y^{2}\right]\right)\equiv-\frac{\partial
H}{\partial R},
\end{array}
\right.
\end{equation}
with
\begin{equation}\label{hfun1}
 H=2D^{2}\frac{R^{\nu+1}}{(\nu+1)}+{\beta}\frac{R^{2(\nu+2)}}{(\nu+2)^{2}}+
 \sigma Y^{2}R^{2(\nu+1)}-2E\frac{R^{\nu+2}}{(\nu+2)}.
\end{equation}
}

By elementary checking one can get convinced that the function $H$
is constant on phase trajectories of both systems (\ref{ds1sp})
and (\ref{ds1h}), and since the integrating multiplier
$\Psi=2R^{\nu},$ occurring in formula (\ref{ds1h}) is positive for
$R>0$, then  phase trajectories of systems (\ref{ds1sp}) and
(\ref{ds1h})  are similar in the right half-plane of the phase
space $(R,\,Y)$. Thus all the statements concerning the geometry
of the  phase trajectories of system (\ref{ds1h}) lying in the
right half-plane are applicable to corresponding  solutions of
system (\ref{ds1sp}). In particular, we immediately conclude that
the critical point $A_{2}(R_{2},\,\,0)$ remains a center when the
nonlinear terms are added. This means that the initial system
(\ref{hdyngen}) possesses a one-parameter family of periodic
solutions. If the right branches of the separatrices of the saddle
$A_{1}(R_{1},\,\,0)$ go to infinity (the stable branch $W^{s}$
when $t\to-\infty$ and the unstable branch $W^{u}$ when
$t\to+\infty$), then the domain of finite periodic motions is
unlimited. Another possibility is connected with the existence of
the homoclinic trajectories.  In this case the initial system
possesses localized soliton-like regimes.
%%16
To answer the question on which of the above mentioned
possibilities is realized in system (\ref{ds1sp}), the behavior
of the saddle separatices, lying to the right from the line
$R=R_1$, should be analyzed.
 We obtain the equation for saddle separatices  by putting
 $H=H(R_{1},0)=H_{1}$ in the LHS of the equation (\ref{hfun1})
 and solving it next with respect to  $Y$:
 \begin{equation}\label{separ1}
Y=\pm \sqrt{\frac{H_{1}+2E{R^{\nu+2}}/{(\nu+2)}-
[2D^{2}\,{R^{\nu+1}}/{(\nu+1)}+{\beta}R^{2(\nu+2)}/{(\nu+2)^{2}}]}
{\sigma R^{2(\nu+1)}}}
\end{equation}

It is evident from equation (\ref{separ1}), that incoming and
outgoing separatrices are symmetrical with respect to $OR$ axis.
Therefore  we can restrict our analysis to one of them, e.g. to
the upper separatrix $Y_{+}$. First of all, let us note, that in
the point$\left(R_1,\,\,0 \right)$ separatrix  $Y_{+}$ forms with
$OR$ axis a positive angle $ \alpha= arctg
\sqrt{{(R_{2}-R_{1})\Psi (R_{1})}/\left({\sigma
R_{1}^{\nu+2}}\right)}. $ The above formula arises from the linear
analysis of system (\ref{ds1sp}) in critical point
$A_{1}(R_{1},\,\,0)$. So  $Y_{+}(R)$ is increasing  when $R-R_{1}$
is small and positive.  On the other hand, function $
G(R)={H_{1}+2E\,{R^{\nu+2}}/{(\nu+2)}-
\left[2D^{2}\,{R^{\nu+1}}/{(\nu+1)}+{\beta}\,R^{2(\nu+2)}/(\nu+2)^{2}\right]},
$ standing inside the square root in equation (\ref{separ1}),
tends to  $-\infty$ as $ R\to +\infty$, because the coefficient
at the highest order monomial $R^{2(\nu+2)}$ is negative, while
the index $\nu+2$ is assumed to be positive.  Therefore the
function $G(R)$ intersects the  open set $R>R_1$ of the $OR$ axis
at least once. Let us denote a point of the first intersection  by
$A_3(R_{3},0)$. The coordinate of intersection satisfies
inequality  $R_{3}
> R_{2}.$ It can be easily seen by noting that $G^{\prime}(R)=
-2R^{\nu}\left({\beta}R^{\nu+3}/{(\nu+2)}-ER+D^{2}\right)=-2R^{\nu}P(R)$
and analyzing $\lim_{R\to R_3^{-}}d\,Y/d\,R$. On the other hand,
$G'(R)<0$ when $R>R_{2}$, and this is suffice to show that
$\lim_{R\to R_3^{-}}d\,Y/d\,R=-\infty.$ Due to the symmetry of the
saddle separatrices, they intersect tangently in the point
$A_3(R_3,\,0),$ The result obtained can be formulated as follows.

%\begin
{\bf Theorem} {\it If $\nu>-2$ and $D^2>\beta\,R_1^{\nu+3}$, then
system (\ref{ds1sp}) possesses a one parameter family of periodic
solutions and the homoclinic solution, formed by the tangent
intersection of separatrices of the saddle point
$A_1\left(R_1,\,\,0 \right)$.
%\end{theorem}
}

\par Thus we have shown, that system (\ref{hdyngen}) possesses
periodic and soliton-like invariant solutions. Let us note in conclusion that
for some special case the integral standing at the RHS of the formula
(\ref{solds1spat}) can be calculated explicitely:
\begin{equation}\label{spatexact}
\omega=\pm\Biggl\{\sqrt{8}\arcsin{\left[\frac{R+1}{2\sqrt{2}}\right]}+\sqrt{2}\,\log{\left[\frac{R-1}
{3-R+ \sqrt{20-{2}(R^2+2R+3)}}\right]}-\omega_0\Biggr\},
\end{equation}
where  $\omega_0=\sqrt{2}\,\pi-\log{2}$. This solution corresponds to the following values of the parameters: $D=1=R_1=\sigma=1,$ $\beta=1/2,$ $\nu=0$, and $E=5/4.$

\section{Conclusions.}

Thus we presented  the effective anzatz  resulting from the
analysis of   the  methods put forward in \cite{fan,baryur,nikbar}
and  based  on the  following observation: the crucial element of
any of anzatz-based method is the closeness of certain class of
functions with respect to the algebraic operations and
differentiating. This observation served as the main motive when
we put forward the anzatz (\ref{vkgenans}). Note, that it can be
easily modified for obtaining the periodic solutions or, more
generally,  exact solutions that can be expressed as algebraic
combinations of some special functions.

The results of the last section  show that the combination of the
self-similar reduction and the qualitative analysis can deliver the
exhaustive answer to the question on the existence of certain
types of solutions within the given family. At the end of this
section we also present the exact homoclinic solution, that
was predicted earlier by the qualitative analysis but failed to be
obtained within the anzatz-based methods. The reason of this is
evidently linked with the fact that solution (\ref{spatexact}) is
too complicated to be expressed as an algebraic combination of the
hyperbolic functions. Whether it can be expressed in terms of
special functions is still an open problem, connected with the
more general problem of finding out a fully universal ansatz.

\end{document}